\documentclass[a4paper]{article}

\usepackage{geometry}
\usepackage{graphicx}
\graphicspath{{fig_gz/}}

\title{The aerogel threshold Cherenkov detector for the
       High Momentum Spectrometer in Hall C at Jefferson Lab}

\author{
R.Asaturyan$^a$, R.Ent$^b$, H.Fenker$^b$, D. Gaskell$^b$, G.M. Huber$^c$, 
M.Jones$^b$, D. Mack$^b$,\\ 
H.Mkrtchyan$^a$, B.Metzger$^b$, N. Novikoff$^d$, V.Tadevosyan$^a$, 
W.Vulcan$^b$, S.Wood$^b$ \\
{\small a.  Yerevan Physics Institute, Yerevan 375036, Armenia }\\ 
{\small b. Thomas Jefferson National Accelerator Facility, Newport News, 
Virginia 23606 USA }\\
{\small c.   Department of Physics, University of Regina, Regina, SK,S4S 0A2, Canada}\\
{\small d.  Houston Baptist University,Houston,Texas 77074 }\\
}
\date{ } 
\begin{document}

\maketitle
\begin{abstract}
We describe a new aerogel threshold Cherenkov detector installed in the HMS 
spectrometer in Hall C at Jefferson 
Lab. The Hall C experimental program in 2003 required an improved particle 
identification system for better identification of $\pi/K/p$, which was 
achieved by installing an additional threshold Cherenkov counter. Two types of
aerogel with $n=1.03$ and $n=1.015$ allow one to reach $\sim\:10^{-3}$
proton and $10^{-2}$ kaon rejection in the 1-5~$GeV/c$ momentum
range with pion detection efficiency better than 99\% (97\%). The detector
response shows no significant position dependence due to
a diffuse light collection technique. The diffusion box was equipped with
16 Photonis XP4572 PMT's. The mean number of
photoelectrons in saturation was $\sim$16 and
$\sim$8, respectively. Moderate particle identification is feasible
near threshold.
\end{abstract}

\section{Introduction}

The Aerogel detector described in this paper was designed and built for 
experiments carried out in Hall C at Jefferson Laboratory. 

A number of $(e,e'h)$ experiments,where a scattered electron is measured in 
coincidence with a hadron,have been performed in Hall C since
1995. The Hall C base experimental equipment consists of two magnetic
spectrometers, the High Momentum Spectrometer (HMS) and the Short Orbit 
Spectrometer (SOS) \cite{CDR}. Depending on the specific requirements of 
experiments, one can detect either negatively-charged (mostly electrons) or
positively-charged particles, by choosing the proper polarity of the magnetic
field and the trigger configuration.

The HMS is designed to detect secondary products of reactions in the momentum 
range $0.5\:GeV/c$ to $7.3\:GeV/c$, while the SOS momentum
extends only up to $\sim\!1.7\:GeV/c$ . Both spectrometers are
equipped with a pair of drift chambers and $X$ - $Y$ timing scintillator
hodoscope planes for trigger formation. 

For particle identification (PID) a combination of Time-of-Flight (TOF), 
threshold gas Cherenkov counter and segmented lead-glass electromagnetic 
calo-rimeter (EC) is used. In addition, for coincidence measurements,
use of the coincidence time difference between scattered electrons and
secondary hadrons is very efficient. But even with perfectly tuned
hodoscope arrays and calibrated detectors, in such a configuration
$\pi/K/p$ separation dramatically deteriorates with momentum as
$\Delta\:t\:\sim\:1/P^{2}$. While TOF is very effective at low
momentum, it becomes in practice useless above $P\:\sim\:3\:GeV/c$.
In addition, in this range hadrons tend to become above the detection threshold
in gas Cherenkov detectors, making $\pi/K/p$ separation more difficult.
Thus, the HMS PID system needed to be augmented for good hadron identification
above $3\:GeV/c$.

A series of Hall C experiments ran in the Summer of 2003 that required
such an improvement of the HMS PID system.
The purpose of the ''Baryon Resonance Electroproduction at High Momentum
Transfer'' \cite{resonance}
experiment was to measure inelastic nucleon transition amplitudes to the
$\Delta (1232)$ and $S_{11} (1536)$ baryon resonances via the 
$p(e,e'p)\pi^{o}$ and $p(e,e'p) \eta$ reactions, respectively, at the 
 previously inaccessible at JLab momentum of
transfer $Q^{2}=7.5\:(GeV/c)^{2}$. The scattered electrons were
detected in the SOS in coincidence with recoil protons of up to $\sim5\:GeV/c$
momentum in the HMS. In this experiment it was important to suppress 
high-momentum pions with respect to protons.

A second experiment, termed ``The Charged Pion Form Factor'',
measured the pion form factor at $Q^{2}=1.6$ and $2.5\: (GeV/c)^{2}$
\cite{Fpi-2}.
In this experiment one detected pions and electrons in coincidence from the
reactions $\gamma_{v} + p \rightarrow \pi^{+} + n$ and
$\gamma_{v} + n \rightarrow \pi^{-} + p$ (in order to estimate contributions from background
 physics processes). Here, the HMS was set up for pion detection. At the highest momentum
setting of this experiment, $P_{HMS}\sim3.4\:GeV/c$, the
ratio of $\pi^{+}$ to protons was expected to be $\sim$1
and good proton rejection became important.

Finally, the experiment ``Duality in Meson Electroproduction{}'' checked
the low-energy cross-section factorization and the quark-hadron duality
phenomenon in semi-inclusive electroproduction of pions (kaons) \cite{mduality}.
Here, it was important to identify kaons and pions at a momentum
$P_{HMS} \ge 3\:GeV/c$. 

The general requirement for these three experiments was a high detection
efficiency for pions in the HMS and the capability to separate protons from 
pions in the first two cases, and pions from kaons in the third case.

The experiments were planned to run at an electron beam intensity up to 90
$\mu$A, hitting a liquid hydrogen (or deuterium) target with length of 4 
cm, rendering rates as high as 1 MHz.

To keep the HMS standard detector configuration intact and not compromise HMS
performance, the new PID detector should be designed with the following 
restrictions:

\begin{enumerate}

\item[-] have large sensitive area to match HMS spectrometer acceptance, with an effective area 
of $\sim$1 m$^{2}$;

\item[-] be slim to fit in $25\:cm$ slot in-between the second drift chamber and
first hodoscope, the only readily available space in HMS detector stack;

\item[-] have minimum material on the particle path to keep the amount of 
multiple scattering and $\delta$-electrons small;

\item[-] have reasonable time resolution and high rate capability.

\end{enumerate}

To obtain a proton threshold momentum of 3-4.6 GeV/c, for Cherenkov radiation, a medium with index
of refraction n = 1.02-1.06 is needed. Aerogel is the best candidate for this
purpose. 
For this reason two types of aerogel material with different indices of
refraction were used.

Many different types of Aerogel detectors have been used in physics 
experiments, but few of them cover the wide acceptance in high intensity beam 
experiments we were looking for. Therefore we designed a new device.

\section{Choice of Aerogel radiators}

The operation of Cherenkov counters is governed by the basic relation
\cite{Groom}
%
%
which connects the emission angle $\theta$ of Cherenkov photons, the velocity
v=$\beta$c of a charged particle and the index of refraction n of the radiator
medium. 
The minimum momentum at which a particle of mass M will exceed the
phase velocity of light in the medium is simply given by
\begin{equation} \label{eq:Pmin}
P_{min} \cdot c = \frac{M \cdot c^{2}}{\sqrt{n^{2}-1}}
\end{equation}
where $c$ is a speed of the light in vacuum.
The number of photons produced by a Z=1 particle per unit track length in a
wavelength region between $\lambda_{1}$ and $\lambda_{2}$
depends on the velocity of the particle and the refractive index n($\lambda$)
of the radiator:
\begin{equation} \label{eq:dNdl}
\frac{dN}{dl} = 2\pi\alpha \left( 1- \frac{1}{n^{2}(\lambda)\beta^{2}} \right)
\left( \frac{1}{\lambda_{1}} - \frac{1}{\lambda _{2}} \right)
\end{equation}
The number of Cherenkov photons scales as $L \cdot sin^{2}\theta$, giving a total of
$N$ detected photons for a radiator of length $L$:
\begin{equation} \label{eq:N}
N=N_{0}L\sin^{2}\theta,
\end{equation}
where

\begin{equation} \label{eq:dNdl}
N_{0} = 2\pi\alpha \left( \frac{1}{\lambda_{1}} - \frac{1}{\lambda _{2}} \right),
\end{equation}
and ${\alpha}$ - is the fine structure constant.

For a diffuse light box having reflectivity M and photodetectors which cover 
an aerial fraction $\epsilon$ of the surface, the average number of detected 
photoelectrons is
\begin{equation} \label{eq:N0}
N_{e} = N_{0} L \left( 1 - \frac{1}{\beta^2 n^2} \right)
\frac{\epsilon}{1-M(1-\epsilon)}.
\end{equation}

Aerogel is a unique material that can have a refractive index between those
typical for gases and liquids (as small as $n=1.008$ and as high as $n=1.094$) \cite{YoYo}.
It is a transparent, highly porous n(SiO$_{2}$)+2n(H$_{2}$O) material with
a density ranging from $\rho$=0.04 to 0.20 g/cm$^{3}$.
The refractive index for the various density aerogel materials is roughly
given by
\begin{equation} \label{eq:refind}
n-1=(0.210\pm0.001)\cdot\rho.
\end{equation}

The optical properties of aerogel can be characterized by an absorption length
$\Lambda_{a}$ and a scattering length $\Lambda_{s}$. A typical value of the
scattering length, at a wavelength $\lambda$ of 400 nm, is
$\Lambda_{s}\sim$2 cm. The absorption length $\Lambda_{a}$ increases almost
linearly in the interval 200-300 nm, and remains nearly constant above that.
At a wavelength $\lambda\sim$400 nm, $\Lambda_{a}\sim$20 cm \cite{optics}.

Taking into account above mentioned requirements of the Hall C experiments,
we chose two different aerogel materials with an
index of refraction of $n=1.030$ and $1.015$, respectively.

The threshold momenta (in GeV/c) for the particles under consideration in these
two types of aerogel materials are presented in the following table:

\vspace{0.5cm}
{\centering \begin{tabular}{|c|c|c|}
\hline 
Type of particle&
P\( _{th} \) in n=1.030&
P\( _{th} \) in n=1.015\\
\hline 
\hline 
$\mu$ & 0.428 & 0.608 \\
\hline
$\pi$ & 0.565 & 0.803 \\
\hline 
K & 2.000 & 2.840 \\
\hline 
P & 3.802 & 5.379 \\
\hline 
\end{tabular}\par}
\vspace{0.3cm}
Table 1. The threshold momenta (P\( _{th} \) in GeV/c) for the different particles in two
type of aerogel.

\begin{figure}
{\par\centering
\includegraphics[clip,trim=10 10 10 10,width=0.75\textwidth]{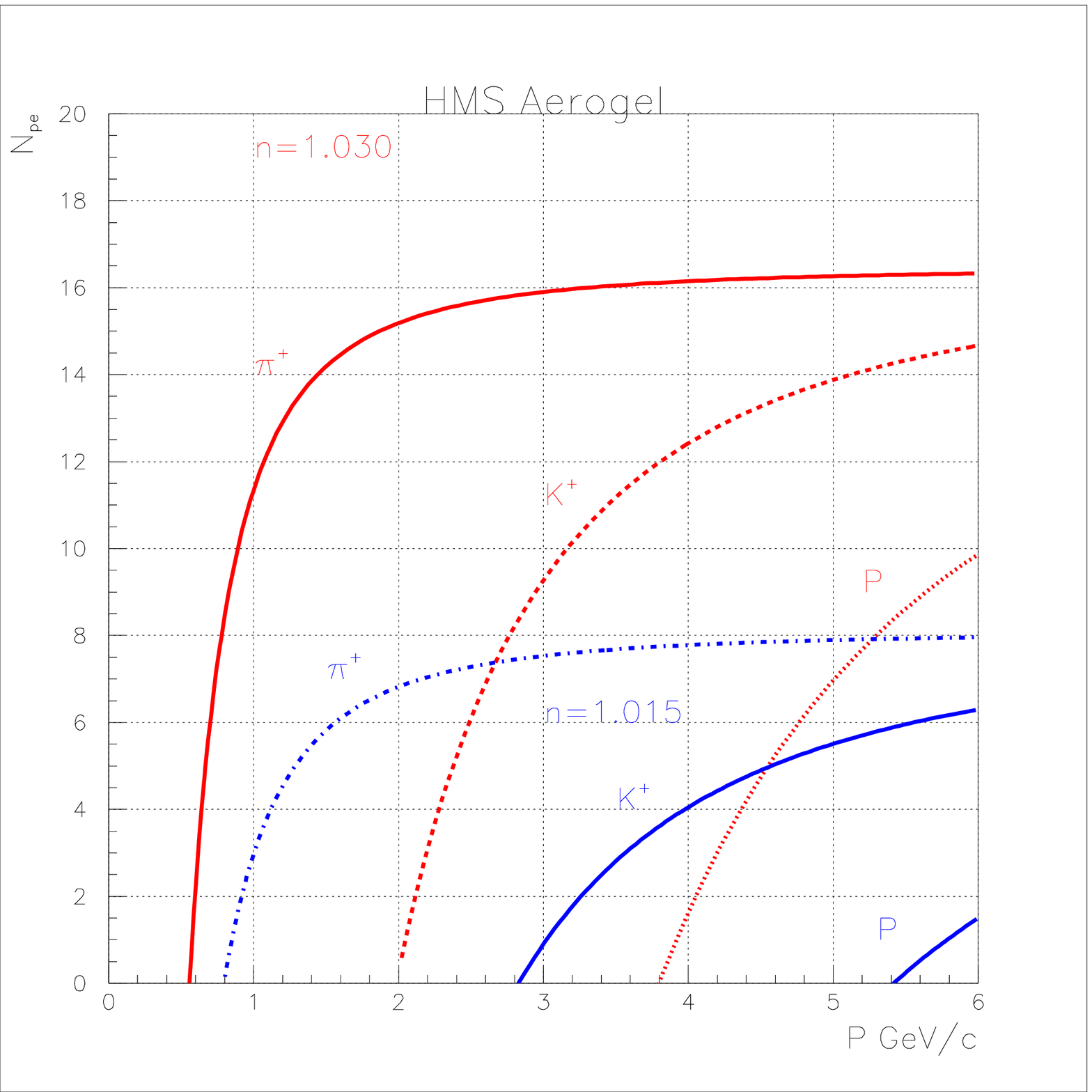} \par}
\caption{Particle separation using Aerogel Cherenkov of different index of
refraction ($N_{0} \sim46$ calculated according to \cite{Higinbotham}). Electrons
(positrons) are above threshold over essentially the full momentum range.}
\label{fig:ramp_up}
\end{figure}

In Fig.~\ref{fig:ramp_up} the expected yields in terms of the number of
photoelectrons are given for both types of aerogel (for a thickness $\sim$9 cm
and a figure of merit $N_{0} \sim46$), as calculated according to \cite{Higinbotham}.

It can be seen that the $n=1.03$ aerogel will allow for good pion/proton
separation up to 4 GeV/c, while the $n=1.015$ aerogel material can be used
for pion/kaon separation in the momentum range 1-3 GeV/c, and 
for pion/proton separation up to 6 GeV/c.

Note that the number of photoelectrons (N$_{pe}$ ) produced just above the
proton (kaon) threshold is much lower for protons (kaons) than for pions,
which allows for limited particle identification just above threshold
by counting N$_{pe}$s.

To obtain the required total thickness and effective area of most
aerogel detectors, they need to be comprised of a large number of the
typically smaller-size tiles. Thus, the uniformity of the optical
quality and the tolerance from tile to tile becomes important.

Aerogels commercially available from Matsushita Electric Works Ltd (Japan)
\cite{matsushita}
are highly transparent and have a light output which is almost linear with the radiator
thickness. They are known to be of high quality.
For example, the detailed study of about 1000 tiles of the Matsushita aerogel
produced for the HERMES experiment \cite{optics} shows that their mean
refractive indexes are in the $n=1.0303 \pm 0.0010$ range, with only small
variation from tile to tile.

The improved light transmittance and hydrophobicity of this ``new'' aerogel
material is due to a new production technique \cite{optics,Adachi}, and makes
it  preferable to the early hydrophilic aerogel materials that
needed baking and storage  in a dry nitrogen atmosphere to maintain the
initial good transmittance of the radiator \cite{Ricks_thesis}.

Although the light transmittance of aerogel is relatively small, the light
absorption is also rather small \cite{optics}. Hence, a large number of photons
undergo multiple scattering and lose directionality but do eventually reach
a photo detector.
Diffuse light collection by means of highly reflective walls, also known
as a ``diffusion box'', seems a good choice.

\section{The Aerogel detector}

\subsection{Physical design}

\begin{figure}
{\par\centering
\includegraphics[clip,width=0.75\textwidth]{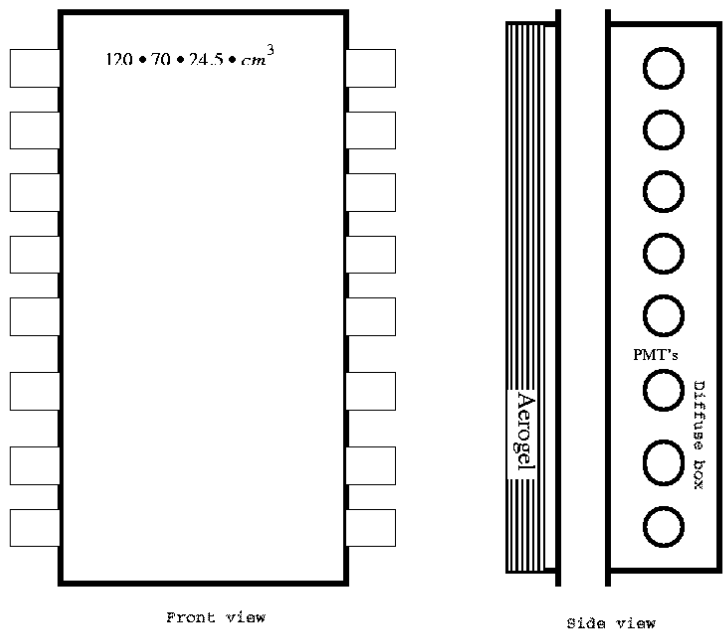} \par}
\caption{Schematic drawing of the Aerogel detector.}
\label{fig:sketch}
\end{figure}

There are different schemes for collecting Cherenkov light. In our detector we
make use of a diffusion box. The photon detection probability in the
case of a diffusion box is directly proportional to the fraction of detector
surface
covered by PMTs.An increase in the area covered by PMTs results in an increase
of the number of photons detected. As a result, and as shown by Monte Carlo calculations, 
we used 16 PMTs in the
counter. 
The aerogel detector schematic design is shown in Fig.~\ref{fig:sketch}.
It is a sandwich of an aerogel tray and a diffusion light box with PMTs. This
allows for simple detector assembly and easy replacement of the aerogel stack.
The active area is 120$\times$70 cm$^{2}$. Eight PMTs are mounted on both
``long'' sides of the box.
The total area covered by the photo-cathode windows of these PMTs
amounts to $\sim$8\% of the inner surface of the counter.
Of course, it is important to have high reflectivity for the
inner walls of the diffusion box.
To accomplish this, the inner walls of the diffusion box were
covered with millipore paper ``Membrane GSWP-0010{}'' (reflectivity 96\%)
\cite{millipore}. 

A Monte Carlo simulation for the HMS aerogel detector was done using a technique to simulate
diffusely reflective aerogel Cherenkov detectors \cite{Higinbotham}.
The detector active area was taken as 120$\times$70$\times$24.5 cm$^{3}$ and 5
inch PMTs with a 20\% quantum efficiency were used as starting point.
Several options were considered for the detector:

\begin{enumerate}

\item[-] readout with PMTs from one and two sides of the diffusion box;

\item[-] the number (5 or 8) of PMTs at each side;

\item[-] two different thicknesses of aerogel, 5 and 9 cm, respectively.

\end{enumerate}

As anticipated, the best uniformity of the summed N$_{pe}$  signal
(flat within $\sim$10\%) was found for the two-side readout.
The mean number of photoelectrons for 5 cm thickness aerogel of $n=1.030$ was
predicted to be $\sim$6.6 for 10 PMT readout and $\sim$8.3
when 16 PMTs (8 PMT's from each side) were used.
As a result of the high optical quality of
the aerogel material under consideration the number of photoelectrons
increased roughly linearly with the aerogel
thickness. The simulated mean number of photoelectrons for the same type of
aerogel of 9 cm thickness was predicted to be $\sim$12.1 for the 10 PMT
readout case and $\sim$14.8 for the 16 PMT case.

We estimated that the use of a detector with $n=1.030$ aerogel material
of 9 cm thickness and with a $\sim$2 mm total Al thickness
for entrance and exit windows would double the number of $\delta$-
electron produced, which could reach $\sim$2\% for HMS detector stack at momentum
$P_{HMS} \geq 4$ GeV/c.

We used 16 ten-stage Photonis XP4572B PMTs of 5 inch diameter, with a
bialkali photo-cathode and a maximum gain of $\sim$10$^{7}$. These PMTs have
a quantum efficiency $\sim$20\% in the wavelength
range $\lambda\sim$350-450 nm, which well matches the transmitted radiation
spectrum of aerogel. Due to an enhanced photoelectron collection efficiency
the effective number of photoelectrons can be increased by a factor of $\sim$2
relative to more commonly used PMTs such as the Burle 8854 \cite{PMTs}.

The close spacing of the metal shields
(to improve the light collection efficiency)
required us to keep the cathode at
ground potential, while positive high voltage was applied to the anode of the
PMT to reduce the
noise level. To compensate for the low gain of the chosen PMT, we modified
their High Voltage (HV) bases by inserting an amplifier
 in the HV dividers as a sequential component after the last
resistor. The fast amplifier was designed \cite{ampl} for standby operation at relatively
low currents with a signal charge amplification factor of $\sim$10. It allows
us to operate the PMTs at lower high voltages, hence prolonging their 
lifetimes.

Two identical boxes were fabricated for the aerogel trays, one each for the
$n=1.030$ and $n=1.015$ materials. Both match the common diffusion box
and can be easily substituted for each other.

Matsushita produces aerogel in the form of
$\sim 110 \times 110 \times 10$ mm$^{3}$ tiles.
In order to stack the material, each tile dimension was first measured and
the differences in block sizes  determined. Taking into account the
tolerances on the actual aerogel material thickness inside the diffusion boxes,
the tiles were layered in 9 stacks in the case of $n=1.030$, and 8 stacks
in the case of $n=1.015$. In both cases the total thickness of aerogel radiator
was 9 cm, using over $\sim$650 tiles for each box.
To prevent any stress on the aerogel material
from the front side of the detector, the aerogel tiles were stacked in a thin
($\sim$5mm) honeycomb sheet and housed in a tray of dimensions 117$\times$67
cm$^{2}$. The layers were shifted relative to each other by about
2-3 cm to prevent any dead zones inside the aerogel volume.

The stacks of aerogel tiles are kept in position by means of a mesh of thin
(100 $\mu$m) stainless steel wire.

\section{Calibration of the photo multiplier tubes}

The calibration of the 16 PMTs consists of evaluating the average number of
detected photoelectrons and distributing them efficiently over the aerogel detector.
The preliminary calibration of each PMT
was performed by measuring the PMT response to a pulsed light source. The light
intensity from the Light Emitting Diode (LED) used was controlled by adjustment
of the height and width of the applied pulses.
For each PMT the Single Photo-Electron peak
position and its width were found versus the applied high voltage. This allowed
us to roughly equalize the response functions for all PMTs, and to determine 
their gains at a given high voltage. 

\begin{figure}
{\par\centering
\includegraphics[clip,trim=10 10 10 10,width=0.75\textwidth]{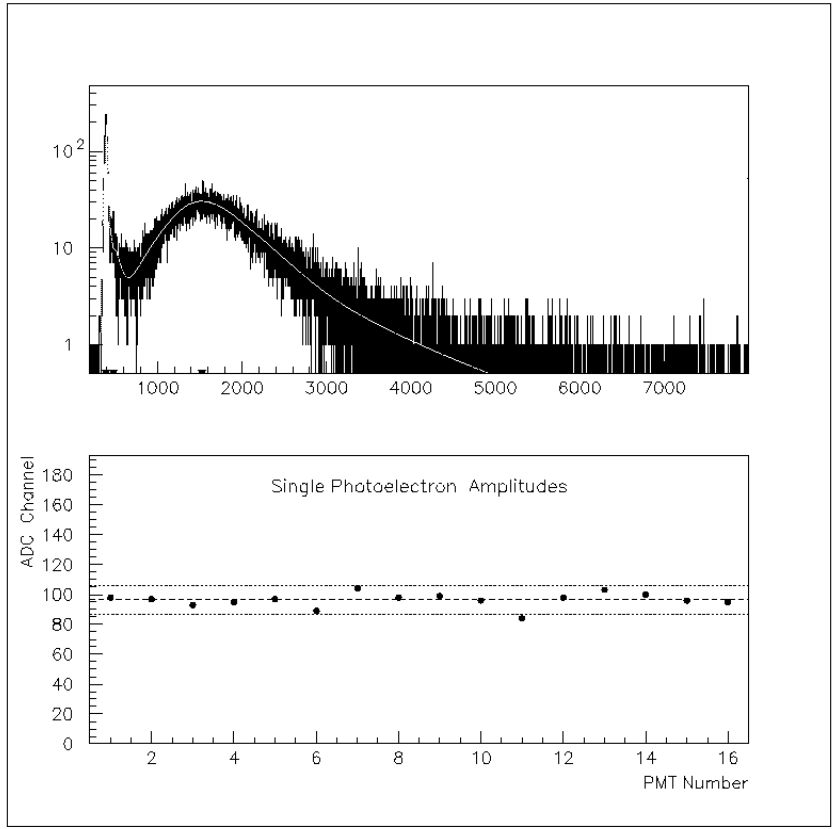}
\par}
\caption{The pulse height distribution of the total sum of the 16 PMTs (top)
and the mean value of the single photo-electron signal for each PMT (bottom),from the cosmic test}
\label{fig:cosm_res}
\end{figure}

A preliminary test of the aerogel detector was performed with cosmic rays. The
detector was positioned horizontally with the diffusion box on top. A pair of
scintillators sandwiched the aerogel detector, with a third one separated by
a layer of lead bricks. This lead absorber was used to select the energy of
cosmic muons (the threshold momentum of muons firing the $n=1.030$ aerogel
detector is $\sim$430 MeV/c). The DAQ system was a simplified
version of the standard DAQ system of Hall C \cite{analysis}.The cosmic test was used 
to roughly adjust the PMT high voltages and estimate
the number of photoelectrons from cosmic muons.
Both a typical pulse height spectrum summed over the 16 tubes, and the
single photo-electron positions (in ADC channels)
for all the tubes after gain matching, are presented in
Fig.~\ref{fig:cosm_res}.

\begin{figure}
{\par\centering
\includegraphics[clip,width=0.75\textwidth]{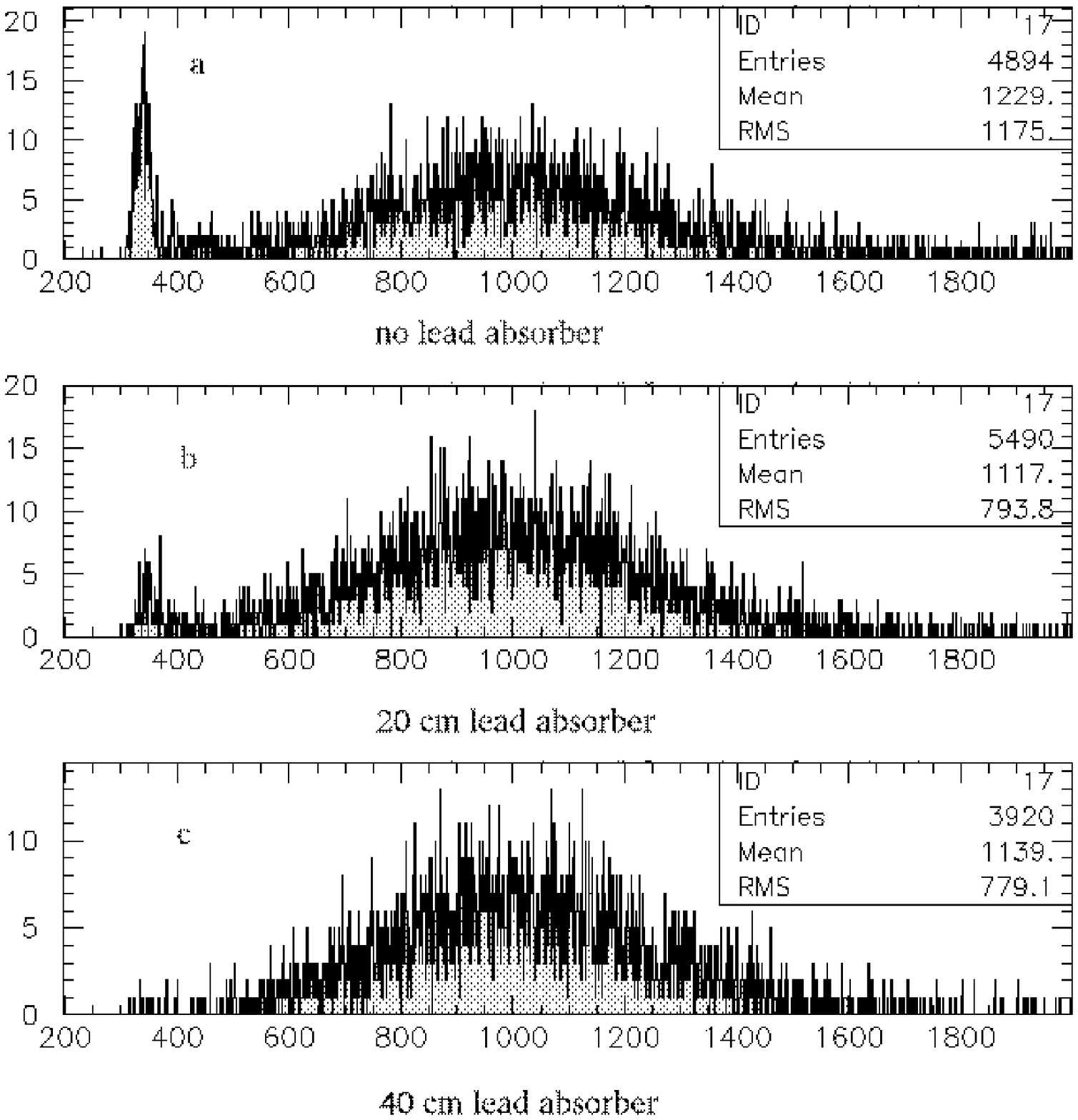} \par}
\caption{The cosmic test results for the aerogel detector with $n=1.030$.
The summed pulse height spectrum is shown for various thicknesses of the
lead absorber used: a) no absorber, b) 20cm lead ($\mu$
momentum \textgreater 300 MeV/c), c) 40 cm lead ($\mu$ momentum P$_{\mu}$
\textgreater 590 MeV/c).}
\label{fig:cosm_res2}
\end{figure}

The ability of the aerogel Cherenkov detector to distinguish between
cosmic muons above and below detection threshold is illustrated
in Fig.~\ref{fig:cosm_res2}, where the pulse height spectrum summed
over the 16 PMTs is shown for several thicknesses of the lead absorber.
The use of the lead absorber between the second and third trigger counters
clearly allows  the low-energy part of the muon spectrum to be cut off.
This is reflected in the figure as the diminishing of the
pedestal events with the increase of the lead absorber thickness.

\section{Experimental results with beam}

The aerogel detector was installed and integrated into the Hall C data
acquisition system, and subsequently successfully used in the
Hall C experimental program of 2003. In this section we will show the results
obtained with the aerogel detector, for both indices of refraction, with beam.

\begin{figure}
{\par\centering
\includegraphics[clip,trim=10 10 10 10,width=0.75\textwidth]{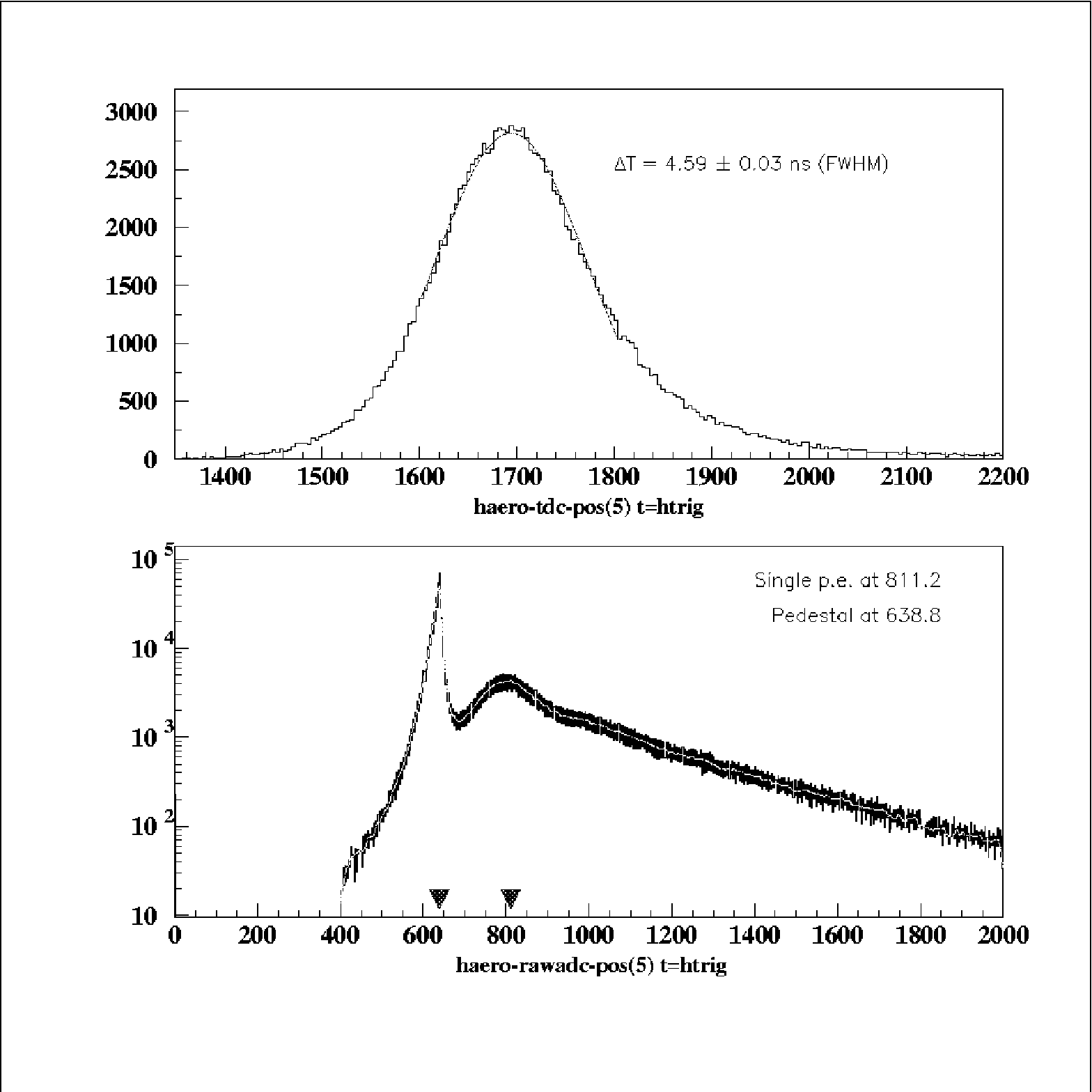}
\par}
\caption{Typical TDC (top) and ADC (bottom) spectra of an Aerogel detector
PMT. The smooth lines are Gaussian (TDC) and spline (ADC) fits to the
histograms shown. The triangles on the bottom of the ADC spectrum indicate
the pedestal and single photo-electron peak positions.}
\label{fig:exp_tdc_adc}
\end{figure}

Fig.~\ref{fig:exp_tdc_adc} shows typical TDC and ADC raw spectra for one
of the aerogel PMTs. Although a $\sim$4 ns (FWHM) time resolution is not 
very good (it is mainly due to the large spread in light paths through the
diffusion box), information from the TDC was still useful in the off-line
analysis for additional rejection of accidental events
in the summed aerogel signal at high rates.
One can see a clean separation of the single photo-electron peak from
pedestals in the ADC spectrum. The two photo-electron peak is also, but
barely, distinguishable. A smooth fit to such ADC spectra provides pedestal
and single photo-electron peak positions for each PMT.
These are then used for detector calibration purposes.

\begin{figure}
{\par\centering
\includegraphics[clip,trim=10 70 10 10,width=0.75\textwidth]{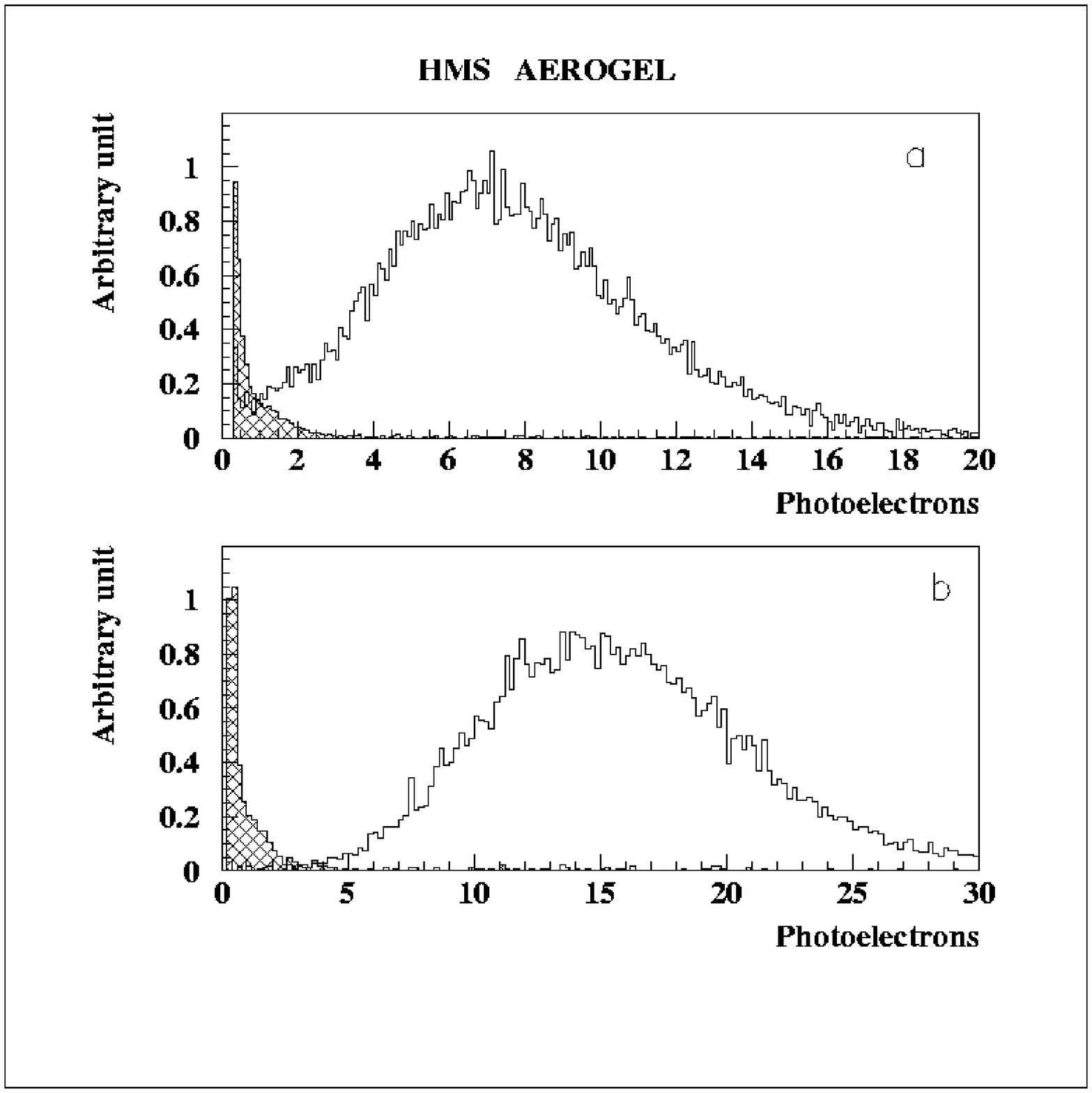} \par}
\caption{The total number of photo-electrons for protons (hatched)
and pions at P$_{HMS}$ = 3.1 GeV/c in aerogel with index of refraction of
 a) $n=1.015$ and b) $n=1.030$.}
\label{fig:tot_Npe}
\end{figure}

In Fig.~\ref{fig:tot_Npe} the distribution of the total number of
photo-electrons N$_{pe}$ (for all PMTs summed) for the aerogel detector
with $n=1.015$ (top) and $n=1.030$ (bottom), respectively, are shown for protons
and pions at 3.1 GeV/c. The signal from the pions is nearly in saturation,
while the signal from protons at this momentum is still below detection
threshold. The mean values of the number of photo-electrons (in saturation)
are $\sim$16 for the $n=1.030$ and $\sim$8 for the $n=1.015$ aerogels,
respectively.

\begin{figure}
{\par\centering
\includegraphics[clip,trim=10 70 10 10,width=0.75\textwidth]{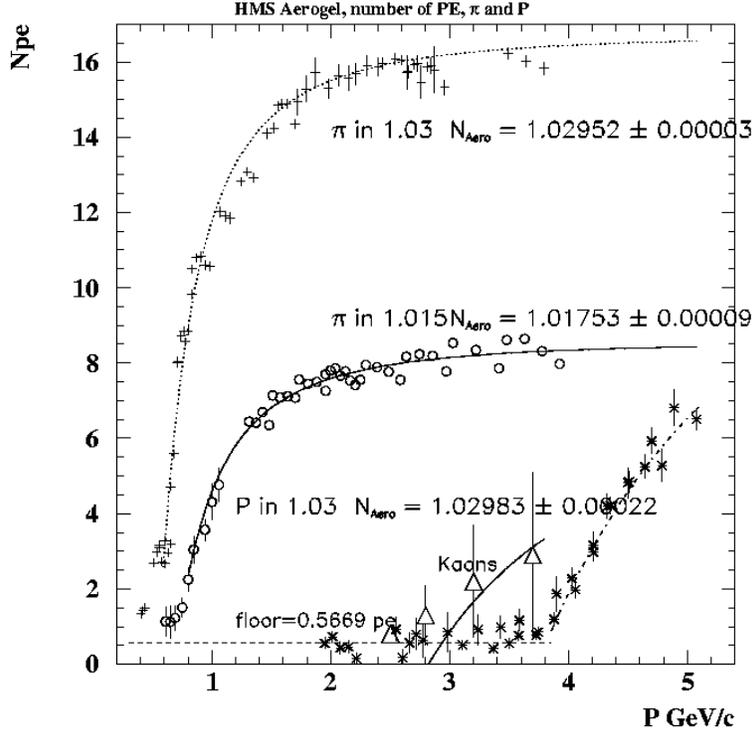}
\par}
\caption{The momentum dependence of N$_{pe}$ for both types of aerogel
material used and for different particles. Both the experimental data
and a fits to them are shown, compare to
Fig.~\ref{fig:ramp_up} }.
\label{fig:exp_ramp_up}
\end{figure}

The experimental data over a wide range of momenta, from 0.5 GeV/c to
4 GeV/c, for different types of particles show that the
dependence of N$_{pe}$ upon momentum has the expected threshold behavior,
and that the number of photo-electrons indeed saturates at high
momentum (see Fig.~\ref{fig:exp_ramp_up}).

When the detector is used in threshold mode, or when one would like
to estimate the threshold velocity of a particle in the given aerogel
(or, alternatively, the index of refraction of the material),
it is important to know the response for particles
below the Cherenkov threshold. The $\sim$0.6 photo-electron background
(for the 16 PMTs summed) shown in Fig.~\ref{fig:exp_ramp_up}
may come from the following sources \cite{delta}:

\begin{enumerate}

\item[-] $\delta$-electrons with momentum above detection threshold;

\item[-] accidental events not rejected by the trigger;

\item[-] particles causing Cherenkov light or scintillation in the
millipore paper or the air in the diffusion box.

\end{enumerate}

After subtraction of this background, one can evaluate the index of
refraction of the used aerogel material from a fit to the data shown in
Fig.~\ref{fig:exp_ramp_up}. The calculated and real values
of index of refraction for both aerogels match well, although better in the
case of the aerogel with $n=1.030$.

\begin{figure}
{\par\centering
\includegraphics[clip,trim=10 70 10 10,width=0.75\textwidth]{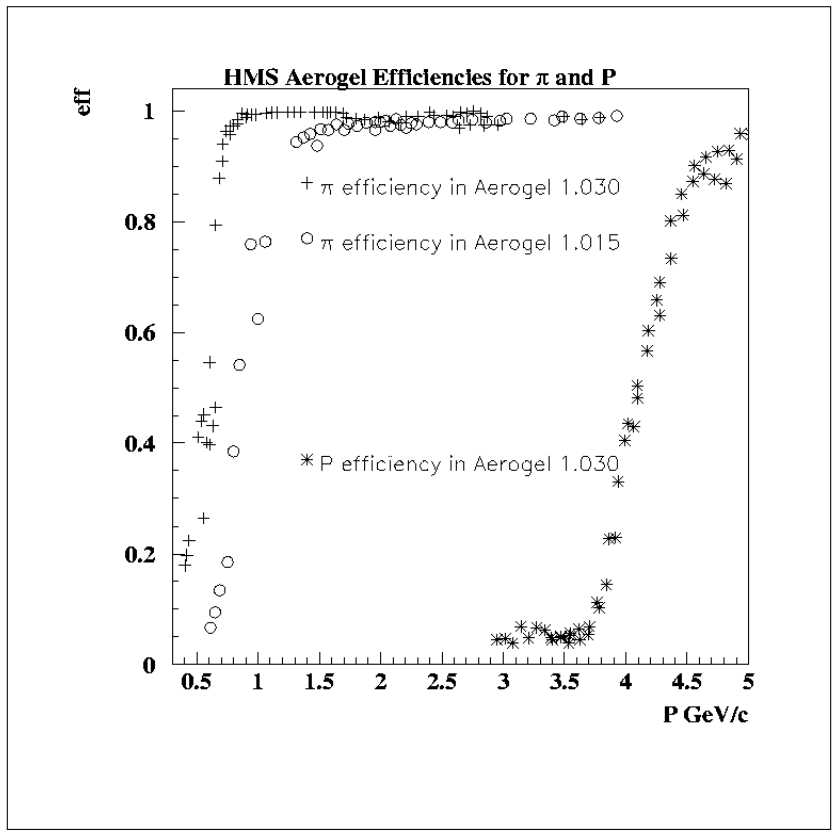}
\par}
\caption{The particle detection efficiency curves for different aerogels
at different levels of cuts.}
\label{fig:exp_eff}
\end{figure}

From these N$_{pe}$ data the detector efficiency versus momentum can be
determined. This results in an efficiency for pion detection in the aerogel
with $n=1.030$ of more than 99\%, in the 1-4 GeV/c momentum range (N$_{pe}\ge 4\:$).
For the case of the aerogel material with $n=1.015$, the pion detection
efficiency is more than 97\%, assuming a cut level of N$_{pe}\sim$2, in a
1.2-4 GeV/c momentum range (see Fig.~\ref{fig:exp_eff}).

\begin{figure}
{\par\centering
\includegraphics[clip,width=0.75\textwidth]{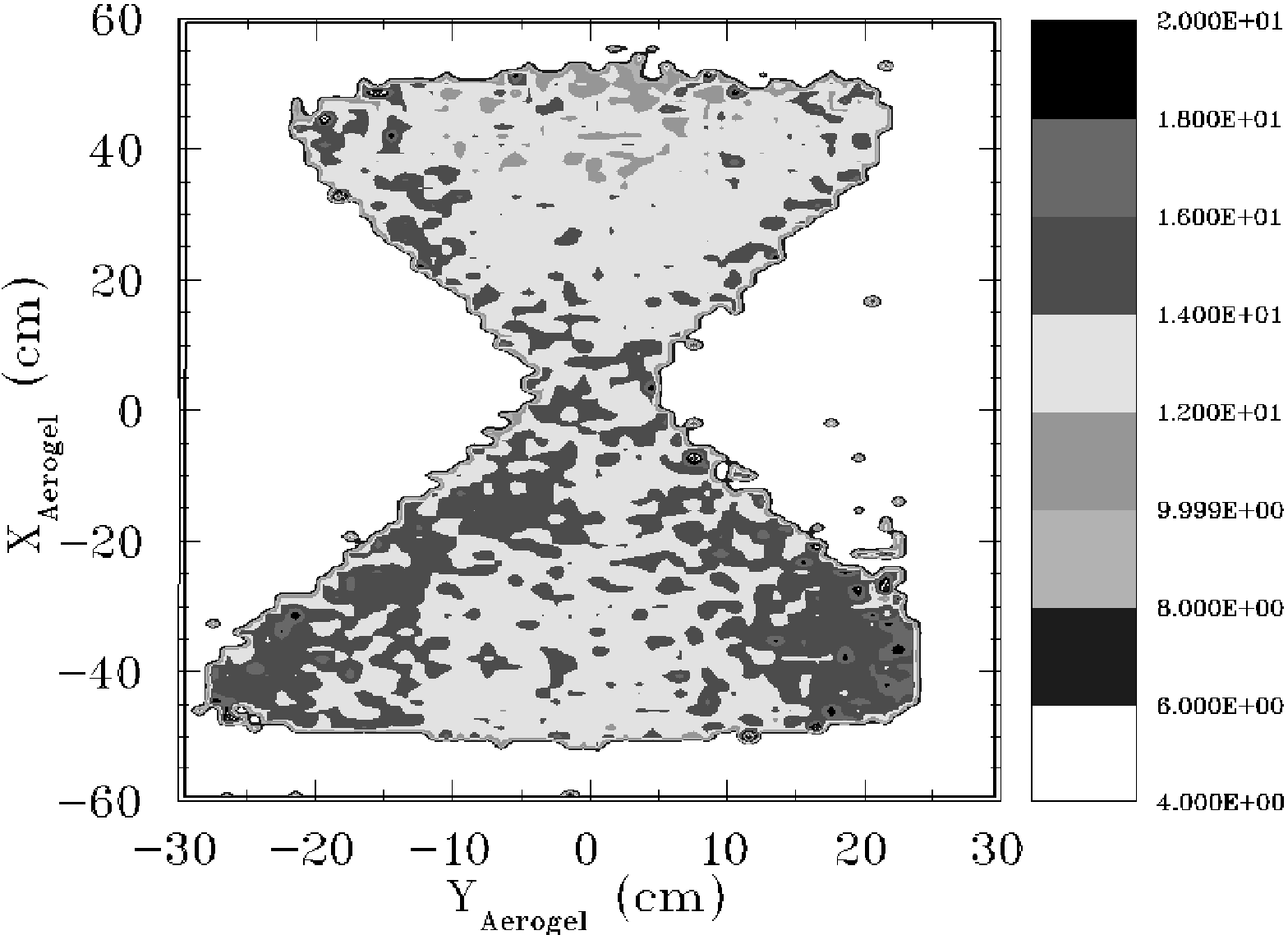} \par}
\caption{The N$_{pe}$ two dimentional distribution on the aerogel detector 
surface  for the ($n=1.030$) aerogel for 3.336 GeV/c pions.}
\label{fig:two_dimen}
\end{figure}

\begin{figure}
{\par\centering
\includegraphics[clip,trim=10 70 10 10,width=0.75\textwidth]{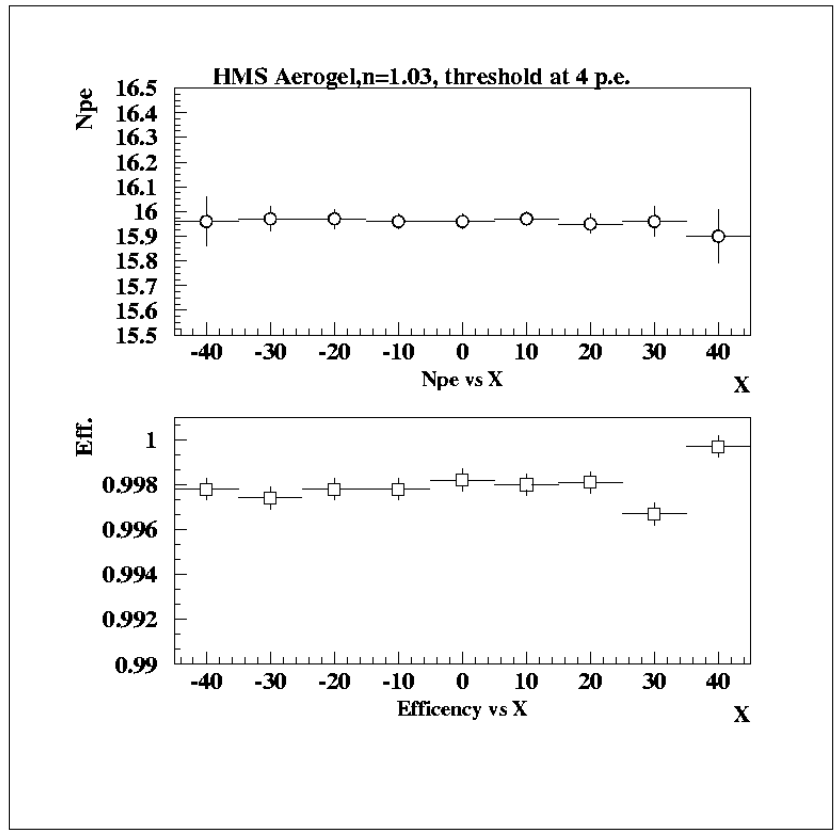} \par}
\caption{The N$_{pe}$ distribution and pion detection efficiency ($P_{\pi}$=
3.2 GeV/c) versus the
vertical X-coordinate for the ($n=1.030$) aerogel detector.}
\label{fig:exp_xdepen}
\end{figure}

\begin{figure}
{\par\centering
\includegraphics[clip,trim=10 70 10 10,width=0.75\textwidth]{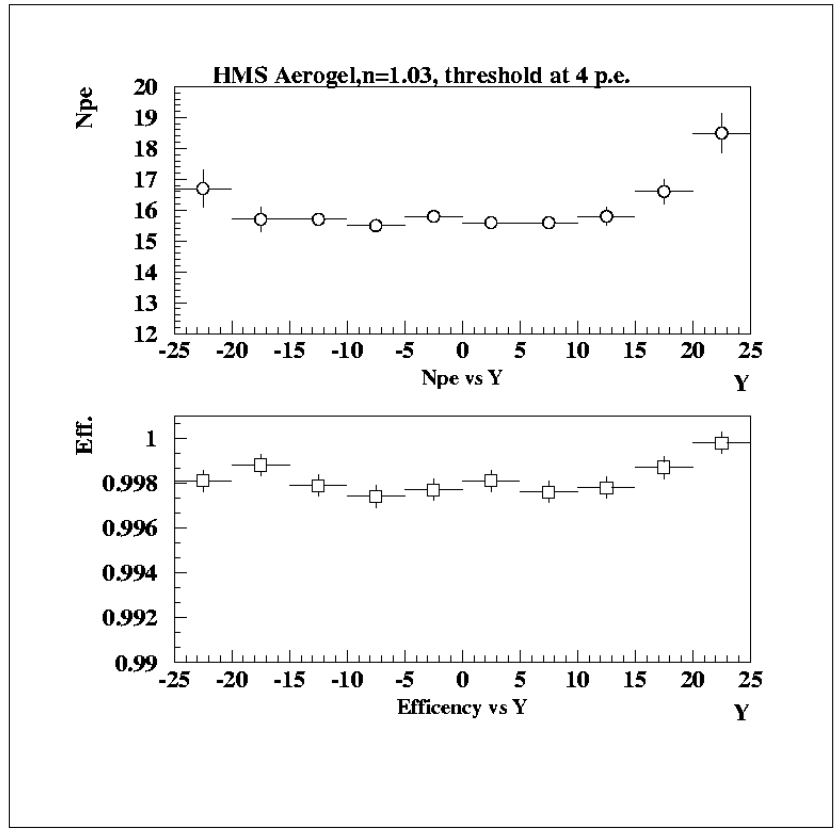} \par}
\caption{The N$_{pe}$ distribution and pion detection efficiency ($P_{\pi}$=
3.2 GeV/c) versus the horizontal Y-coordinate for the ($n=1.030$) aerogel 
detector. The two sets of PMTs are located at Y=${\pm}$25 cm. }
\label{fig:exp_ydepen}
\end{figure}

\begin{figure}
{\par\centering
\includegraphics[clip,trim=10 70 10 10,width=0.75\textwidth]{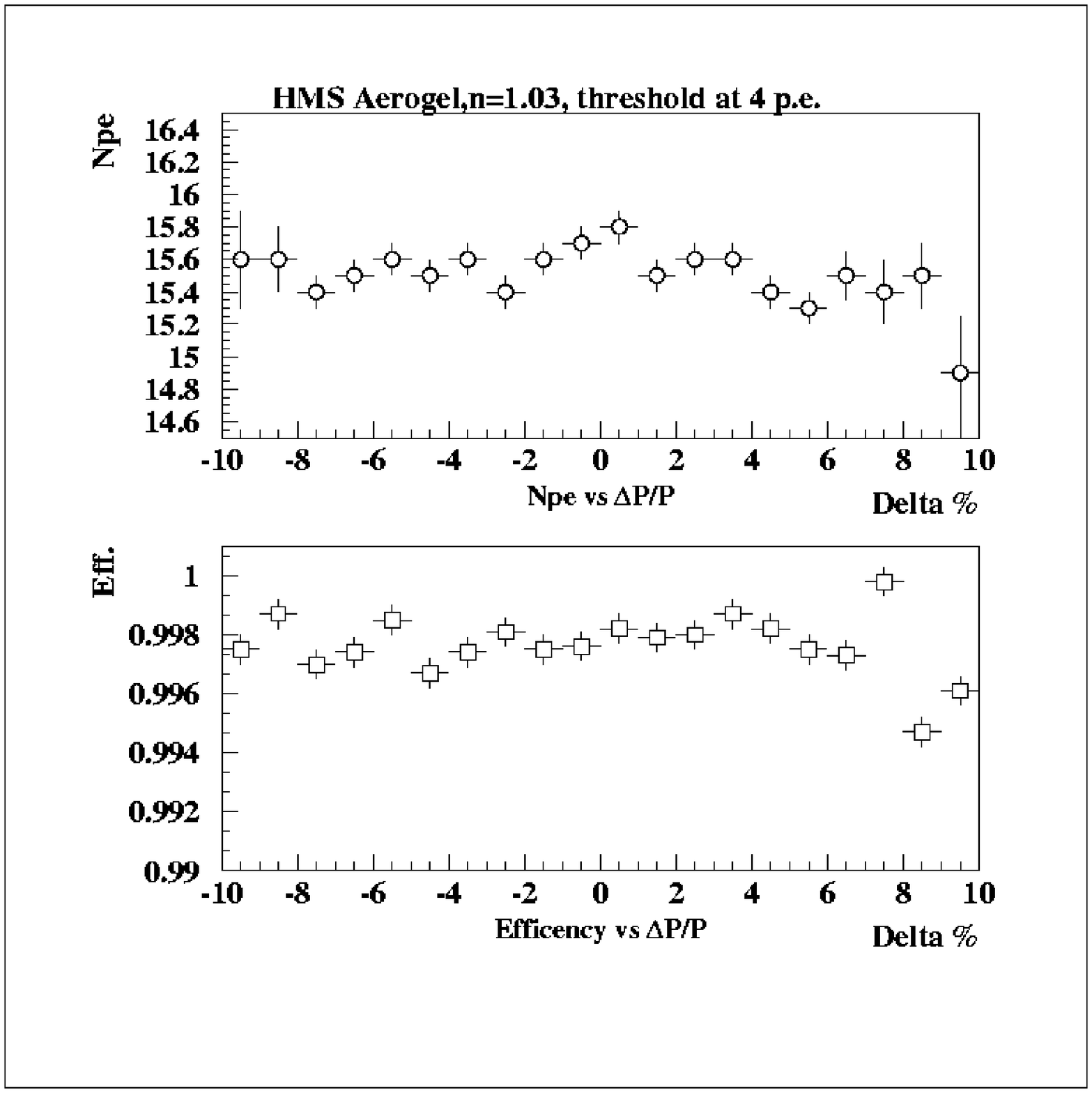}
\par}
\caption{The N$_{pe}$ distribution and pion detection efficiency versus the HMS
fractional momentum for the ($n=1.030$) aerogel detector.}
\label{fig:exp_deltadep}
\end{figure}

One of the most important features of any detector is the near independence
of its response function to position.As shown in Figs.~\ref{fig:two_dimen} ,~\ref{fig:exp_xdepen}
and ~\ref{fig:exp_ydepen}, the total sum of photo-electrons detected by
the aerogel detector has a near flat distribution both in the vertical (X)
and horizontal (Y) direction .
Not surprisingly, close to the PMTs some
enhancement can be seen in the number of photo-electrons detected.
Similarly, in Fig.~\ref{fig:exp_deltadep} the response function versus the
spectrometer fractional momentum ($|{\Delta}p/p| \le $10\%) is shown.
There is no significant dependence over the full momentum acceptance of the  HMS.

Note that for these last cases a cut of $N_{pe}>4$ was used to provide more
than 99\% detection efficiency for the ($n=1.030$) aerogel detector.
The aerogel material with index of refraction ($n=1.015$) shows a similar
behavior, but with less detection efficiency (97\% for a $N_{pe}>2$ cut
condition).

The long-term stability of the HMS aerogel detector was tested during
experimental runs in Hall C over a 6-month period. The mean number of
photo-electrons for the summed detector signals remained stable to
within $\sim$2\%, with a typical particle rate of $\sim$500-600 kHz.

Our studies show that, for an aerogel detector with ($n=1.015$) where kaons
cross detection threshold at a momentum of 2.8 GeV/c, it is more efficient
to use the aerogel detector to reject kaons than to select them.
The number of photo-electrons generated by these kaons is about 4,
for a momentum up to 4 GeV/c, as shown in Fig.~\ref{fig:ramp_up} and
Fig.~\ref{fig:exp_ramp_up}. Applying a cut in N$_{pe}$ of $\sim$2 rejects 
$\sim 95 \% $ of kaons at a momentum P = 3.7 GeV/c, while the same cut rejects
only $\sim 6 \% $ of kaons at a momentum of 2.4 GeV/c (below threshold).

\section{Conclusions}

The particle identification properties of the HMS spectrometer in Hall C at
Jefferson Lab have been significantly improved by adding
a flexible aerogel threshold Cherenkov detector. The detector consists of
an aerogel material followed by a light diffusion box. The radiator tray can
easily be swapped for an alternate one with aerogel material with different
index of refraction. The addition of this detector enhanced the
capabilities of the spectrometer in distinguishing protons from pions
on the level of $2.8-1.1\cdot10^{-3}$ (for aerogel with $n=1.030$) with a pion
detection efficiency better than 99\% in the 1-4 GeV/c momentum range.
It allowed the distinction of kaons from pions on the level of
$10^{-2}$, for aerogel with $n=1.015$, with a pion detection efficiency better
than 97\% in a 1.2-4 GeV/c momentum range.
The mean numbers of detected photo-electrons are $\sim$16 and $\sim$8 for
the $n=1.030$ and $n=1.015$ aerogel material, respectively. The detector response
is uniform to within $\sim$10\% over the full effective area. The experimental
results are in good agreement with expected values from simulations using a
standard Monte Carlo program for aerogel detectors \cite{Higinbotham}.
The number of fake photo-electrons for particles below detection threshold
reaches $\sim$0.6, which may be a result of  $\delta$-electrons,accidental events
or scintilations of particles traversing the detector.

We wish to thank many people who assisted and contributed in the design,
construction and testing of this detector. We are particularly indebted to
R.~Carlini for support to construct such a detector, D.~Higinbotham for
assistance and providing access to his Monte Carlo simulation program
for Aerogel detectors, B.~Wojtsekhowski for interest and many useful
discussions, C.~Zorn for valuable contributions to the systematic tests of
the PMTs, V.~Popov for the development of the PMT amplifier and the
installation of these in the HV bases, W.~Kellner and his group for their
technical expertise and help with the installation of the aerogel detector
and the aerogel radiator replacement in Hall C. Lastly, many thanks to
J.~Beaufait for continuous help during all stages of the construction and
the preliminary testing.

The Southeastern Universities Research Association (SURA) operates the Thomas
Jefferson National Accelerator Facility for the United States Department of
Energy under contract DE-AC05-84ER40150.

\end{document}